\documentclass[prd,preprint,tightenlines,%
superscriptaddress,showpacs,floatfix,nofootinbib
]{revtex4}
\usepackage{epsfig}
\begin{document}
\title{How to define physical properties of unstable particles}
\author{J.~Gegelia}
\affiliation{Institut f\"ur Kernphysik, Johannes
Gutenberg-Universit\"at,  D-55099 Mainz, Germany}
\affiliation{High Energy Physics Institute, Tbilisi State University, Tbilisi,
Georgia}
\author{S.~Scherer}
\affiliation{Institut f\"ur Kernphysik, Johannes
Gutenberg-Universit\"at, D-55099 Mainz, Germany}
\date{October 22, 2009}

\begin{abstract}
   In the framework of effective quantum field theory we address
the definition of physical quantities characterizing unstable particles.
   With the aid of a one-loop calculation, we study this issue in terms of
the charge and the magnetic moment of a spin-1/2 resonance.
   By appealing to the invariance of physical observables under field
redefinitions we demonstrate that physical properties of unstable particles should
be extracted from the residues at complex (double) poles of the corresponding
$S$-matrix.
\end{abstract}

\maketitle
\section{Introduction}
   The question how to define a theory of unstable particles
which is consistent with general requirements of a relativistic
quantum field theory has a long history (see, e.g., Refs.
\cite{Matthews:1958sc,Matthews:1959sy,Jacob:1961zz} for early work).
   In the beginning, the discussion primarily focussed on the definition
of the mean mass and the mean lifetime as static characteristics of an
unstable particle.
   In the early 1990s, this issue attracted considerable renewed attention in
the context of the Standard Model.
   For instance, in Refs.~\cite{Willenbrock:1990et,Valencia:1990jp} an example was given
in the scalar sector showing the field-redefinition dependence of the mass once defined as the
zero of the real part of the inverse propagator.
   Such a definition corresponds to a relativistic Breit-Wigner mass parameter.
   Furthermore, for the $Z$ boson it was shown in Refs.\
\cite{Sirlin:1991fd,Sirlin:1991rt,Willenbrock:1991hu,Gegelia:1992kj,Gambino:1999ai}
that, at two-loop order, the Breit-Wigner mass is gauge-parameter
dependent.
   In contrast, the mass of an unstable particle defined through
the real part of the pole of the propagator is field-redefinition
and gauge-parameter independent \cite{'tHooft:1972ue,Lee:1972fj,Balian:1976vq}
and therefore qualifies as a physical quantity.

   As there is no fundamental dynamical theory of hadron resonances the
problem of field-redefinition invariance and gauge-parameter
(in)dependence received little attention for these unstable
particles \cite{Willenbrock:1991hu}.
   The characteristic properties of hadron resonances eventually have to
be described by QCD.
   With the progress of lattice techniques
\cite{Wittig:2008zz,Alexandrou:2008bn,Braun:2009jy,Boinepalli:2009sq}
and, especially, the low-energy effective theories (EFT) of QCD (see, e.g.,
\cite{Weinberg:1979kz,Gasser:1983yg,Gasser:1988rb,Scherer:2002tk,Pascalutsa:2006up,Scherer:2009bt}
and references therein) it is timely to reinvestigate the definition of
physical characteristics of resonances.
   In the present work we study this issue in terms of the
charge and the magnetic moment of a spin-1/2 resonance.
   The choice of electromagnetic properties is motivated by the
fact that there already exists an extensive experimental and theoretical
program for investigating resonance photon decay amplitudes and nucleon resonance
transition form factors (see, e.g., \cite{Tiator:2003uu,Burkert:2004sk,Capstick:2007tv,Burkert:2008mw}).
   Moreover, experiments aiming at the extraction of magnetic moments of
excited baryons have already been performed or are planned
(see, e.g., Ref.\ \cite{Kotulla:2008zz} for an overview).

\section{The model and definitions}
   In order to keep the technicalities as simple as possible, while at
the same time studying a sufficiently non-trivial physics case,
we consider a toy model for a positively charged, unstable heavy fermion ($\Psi$)
(``resonance'') which may decay into a positively charged, stable light fermion $(\psi)$
(``proton'') and a stable, neutral pseudoscalar $\phi$ (``neutral pion'').
   We make use of the following effective
Lagrangian,
\begin{eqnarray}
\mathcal{L}_{\rm eff} & = & \bar{\psi}\,(i
D\hspace{-.65em}/\hspace{.1em}-M_N)\psi + \bar\Psi(i
D\hspace{-.65em}/\hspace{.1em}-M_R)\Psi
\nonumber\\
&& -\frac{1}{4}\,F^{\mu\nu}F_{\mu\nu}
+\frac{1}{2}\,\partial_\mu\phi
\partial^\mu\phi-\frac{M^2}{2}\,\phi^2 \nonumber\\
&& -i g\,\phi \left(\bar\psi\,\gamma^5\Psi+\bar\Psi\,\gamma^5 \psi\right)
\nonumber\\
&& -e \,\kappa\,F_{\mu\nu}\,\bar\Psi\,\sigma^{\mu\nu}\Psi
+ \cdots \,, \label{LfreiA}
\end{eqnarray}
where $D^\mu=\partial^\mu+i e A^\mu$ ($e>0$) denotes the covariant
derivative acting on the positively charged fields and
$F^{\mu\nu}=\partial^\mu A^\nu-\partial^\nu A^\mu$ is the usual
electromagnetic field strength tensor.
   For the decay interaction we take a simple pseudoscalar coupling with
a real coupling constant
$g$.\footnote{For the present purposes the details of the interaction
allowing the resonance to decay into the nucleon and the pseudoscalar are
not important. We could as well have chosen a derivative interaction.}
   For illustrative purposes, we also include a coupling of
$F_{\mu\nu}$ to the resonance with strength $e\kappa$,
where $\sigma^{\mu\nu}=\frac{i}{2}[\gamma^\mu,\gamma^\nu]$.
   The ellipses stand for an infinite number of interaction
terms respecting Lorentz invariance and the discrete symmetries
$C$, $P$, and $T$.
   Throughout this paper we will make use of dimensional regularization with $n$
space-time dimensions.
   We do not show any counter terms explicitly but rather subtract the divergences
of loop diagrams using the $\overline{\rm MS}$ scheme.
   The loop integrals appearing in the final expressions of our calculation are
therefore to be understood as $\overline{\rm MS}$ subtracted.

   The following discussion will rely on the fact that physical quantities
should remain invariant under a field transformation
\cite{Kamefuchi:1961sb,Coleman:1969sm,Scherer:1994aq,Weinberg:1995mt,Fearing:1999fw}.
   Let us perform in the Lagrangian of Eq.~(\ref{LfreiA})
the field transformation
\begin{equation}
\Psi\mapsto \Psi+i\xi \,\phi\,\gamma^5\psi\,, \  \ \bar\Psi\mapsto \bar\Psi
+i \xi\,\phi\,\bar\psi\,\gamma^5 , \label{ftr}
\end{equation}
where $\xi$ is an arbitrary real parameter with the dimension of an inverse mass.
   This transformation generates
\begin{eqnarray}
\Delta \mathcal{L}& = & -i\xi\,M_R\,\phi \left(
\bar\psi\,\gamma^5\Psi+\bar\Psi\,\gamma^5\psi\right)\nonumber\\
&& -i\xi e \phi A_\mu \left( \bar\psi\,\gamma^5
\gamma^\mu\Psi+\bar\Psi\gamma^\mu\,\gamma^5 \psi\right)\nonumber\\
&& -i\,\xi\,e\,\kappa\,F_{\mu\nu}\,\phi
\left(\bar\psi\,\gamma^5\,\sigma^{\mu\nu}\Psi+\bar\Psi\,
\sigma^{\mu\nu}\,\gamma^5\psi\right)
\nonumber\\
&&- \xi\phi\left( \bar\psi\,\gamma^5\gamma^\mu
\partial_\mu\Psi+\bar\Psi\gamma^\mu
\,\gamma^5\partial_\mu \psi\right)\nonumber\\
&& -\xi \partial_\mu\phi \bar\Psi\gamma^\mu\,\gamma^5\psi
+ 2\,\xi\,g\,\phi^2 \bar\psi\,\psi+\cdots\,, \label{NewIntTerms}
\end{eqnarray}
where we have only displayed the terms linear in $\xi$ which originate from
the expression explicitly shown in Eq.~(\ref{LfreiA}).
\begin{figure}
\epsfig{file=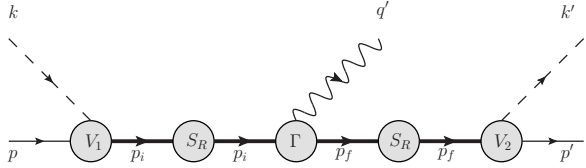,width=0.5\textwidth}
\caption[]{\label{UGProcess:fig} Resonant part of the process
$\phi\,\psi\to \gamma\,\phi\,\psi $.
   The thick, thin, dashed, and wiggly lines correspond to
the heavy fermion, the light fermion, the neutral pseudoscalar, and the
photon, respectively.}
\end{figure}

   Our aim is to investigate the charge and the magnetic moment of the heavy
fermion.
   As this fermion is an unstable ''particle'' it cannot
appear in an asymptotic state.
   Therefore, let us consider the amplitude of the process
$$\phi(k)+\psi(p)\to \gamma(q')+\phi(k')+\psi(p') $$ for an invariant
energy near the mass of the resonance
(see Fig.~\ref{UGProcess:fig}).
   The total amplitude $A$ will be the sum of a resonant part $A_{\rm r}$ and
a non-resonant part $A_{\rm n.r.}$.
   The resonant part can be written as
\begin{eqnarray}
A_{\rm r} & = & -i\,e\,\epsilon_\mu^\ast\, V_2(p',p_f) i S_R(p_f) \Gamma^\mu(p_f,p_i) i
S_R(p_i) V_1(p_i,p)\,, \label{respart}
\end{eqnarray}
where $\epsilon$ denotes the polarization vector of the photon, and
$p_i=p+k$ and $p_f=p'+k'$ refer to the intermediate resonance momenta before
and after the electromagnetic interaction, respectively.
  The dressed propagator of the heavy fermion $S_R$ reads
\begin{equation}
 S_R(p) = \frac{1}{p\hspace{-.45 em}/\hspace{.1em}-M_R
-\Sigma(p\hspace{-.45 em}/\hspace{.1em})}\,,\label{dressedDpr}
\end{equation}
where $-i\Sigma (p\hspace{-.45 em}/\hspace{.1em})$ is the sum of the
one-particle-irreducible diagrams contributing to the two-point
function of the heavy fermion.

   The dressed propagator $S_R$ has a complex pole which is obtained
by solving the equation
\begin{equation}
z - M_R -\Sigma(z)=0\,. \label{poleequation}
\end{equation}
   We define the pole mass as the real part of $z$.
   In the vicinity of the pole, $S_R$ can be written as
\begin{equation}
S_R(p)= \frac{Z}{p\hspace{-.45 em}/\hspace{.1em}-z}+{\rm n.p.}=
\frac{Z \left( p\hspace{-.45 em}/\hspace{.1em} +
z\right)}{p^2-z^2}+{\rm n.p.} \,, \label{SRparametrization}
\end{equation}
where n.p.~generically denotes non-pole, i.e.~regular terms.
   The residue $Z$ is given by
\begin{equation}
Z=1+\delta Z = \frac{1}{1-\Sigma'(z)}\,.
\label{wfrc}
\end{equation}

   For purely technical convenience in the calculations to follow, let us
introduce ``Dirac spinors'' with complex masses $z$,\footnote{Note that
$\bar w^i(p)\neq w^{i\dagger}(p)\gamma_0$.}
\begin{eqnarray}
w^1(p) & \equiv & \sqrt{p_0+z} \left(
\begin{array}{c}
 1 \\
 0 \\
 \frac{p_z}{p_0+z} \\
 \frac{p_x+i p_y}{p_0+z}
\end{array}
\right)\,,\nonumber\\
\bar w^1(p) & \equiv & \sqrt{p_0+z}\left(
\begin{array}{llll}
1\,, & 0\,, & \frac{-p_z}{p_0+z}\,, &
-\frac{p_x-i p_y}{p_0+z}
\end{array}
\right) ,\nonumber\\
w^2(p) & \equiv & \sqrt{p_0+z}
 \left(
\begin{array}{c}
 0 \\
 1 \\
 \frac{p_x-i p_y}{p_0+z} \\
\frac{- p_z}{p_0+z}
\end{array}
\right)\,,\nonumber\\
\bar w^2(p) & \equiv &
\sqrt{p_0+z}
\left(
\begin{array}{llll}
0\,, & 1\,, & -\frac{p_x+i p_y}{p_0+z}\,, &
\frac{p_z}{p_0+z}
\end{array}
\right). \label{spinorsdef}
\end{eqnarray}
   For $p^2=z^2$, the spinors satisfy the ``Dirac equations''
\begin{eqnarray}
\left( p\hspace{-.45 em}/\hspace{.1em}-z\right)w^i&=&0\,,\nonumber\\
\bar w^i \left( p\hspace{-.45 em}/\hspace{.1em}-z\right)&=&0
\label{complexDEquations}
\end{eqnarray}
as well as the identity
\begin{equation}
w^i(p)\,\bar w^i(p) =  p\hspace{-.45 em}/\hspace{.1em}+z\,,
\label{projector}
\end{equation}
where a summation over $i=1,2$ is implied.
   For $p^2\neq z^2$, the difference between $w^i(p)\,\bar w^i(p)$  and
$p\hspace{-.45 em}/\hspace{.1em}+z$ is of ${\cal O}(p^2-z^2)$ so that
we can write for the dressed propagator
\begin{equation}
S_R(p) = \frac{Z \, w^i(p)\,\bar w^i(p)}{p^2-z^2}+{\rm n.p.} \,.
\label{SRRewritten}
\end{equation}
   Substituting Eq.~(\ref{SRRewritten}) in Eq.~(\ref{respart}) we
decompose the resonant amplitude in a double-pole contribution $A_{\rm d.p.}$ and
the rest \cite{Sirlin:1991fd}, where
\begin{eqnarray}
A_{\rm d.p.} & = &  -i\,e\,\epsilon_\mu^\ast\, V_2(p',p_f) w^i (p_f) \sqrt{Z} \frac{i}{p_f^2-z^2}
\nonumber\\
&&\times \sqrt{Z}\,\bar w^i(p_f) \Gamma^\mu(p_f,p_i)
w^j(p_i)\sqrt{Z}\nonumber\\
&&\times \frac{i}{p_i^2-z^2} \sqrt{Z} \bar w^j(p_i) V_1(p_i,p)\,. \label{polepart}
\end{eqnarray}
Using Eqs.~(\ref{complexDEquations}), we parameterize the
renormalized vertex function for $p_f^2=p_i^2=z^2$
in terms of two form factors,
\begin{equation}
\sqrt{Z}\,\bar w^i(p_f) \Gamma^\mu(p_f,p_i)
w^j(p_i)\sqrt{Z}=
\bar w^i(p_f)\left[\,\gamma^\mu\, F_1 (q^2)  +
i \,\sigma^{\mu\nu}\, q_\nu\,F_2(q^2)\,\right]w^j(p_i) \,, \label{FF}
\end{equation}
where $q=p_f-p_i$.
   Note that our normalization of $F_2$ differs by a constant factor
from the one commonly used for stable particles.

\section{Form factors}
\begin{figure}
\epsfig{file=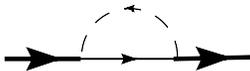,width=0.2\textwidth}
\caption[]{\label{DeltaMassInd:fig} One-loop self-energy diagram
of the heavy fermion. The thick, thin, and dashed lines correspond to
the heavy fermion, the light fermion, and the neutral pseudoscalar, respectively.}
\end{figure}

\begin{figure}
\epsfig{file=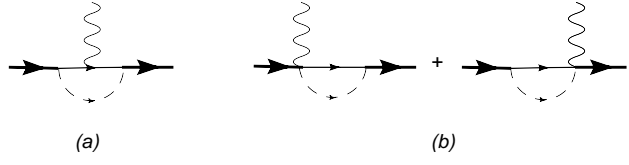,width=0.5\textwidth}
\caption[]{\label{VertexFunction:fig} One-loop diagrams
contributing to the photon heavy-fermion vertex function.
The thick, thin, dashed, and wiggly lines correspond to
the heavy fermion, the light fermion, the neutral pseudoscalar,
and the photon, respectively.}
\end{figure}

   Below, we calculate the form factors $F_1$ and $F_2$ to one-loop order
at $q^2=0$.
   For that purpose we need to obtain the one-loop contributions
to the wave function renormalization constant $Z$ and the vertex
function $\Gamma^\mu$ at $q^2=0$.
   The one-loop contribution to the heavy-fermion self-energy
generated by the interaction terms explicitly given in
Eqs.~(\ref{LfreiA}) and (\ref{NewIntTerms}) is shown in
Fig.~\ref{DeltaMassInd:fig}.
   We obtain for $\delta Z$ of the residue $Z=1+\delta Z$ of Eq.\ (\ref{wfrc}),
\begin{equation}
\label{deltaZ}
\delta Z=g^2 \delta Z_a+g\xi\delta Z_b\,,
\end{equation}
where the explicit expressions for the
coefficients $\delta Z_a$ and $\delta Z_b$ are given in Eqs.~(\ref{deltaZa})
and (\ref{deltaZb}) in the appendix.
   The contribution of the tree-order vertex diagram to $\Gamma^\mu$ reads
\begin{equation}
\Gamma_{\rm tree}^\mu = \gamma ^{\mu }
+2\kappa\, i \,\sigma^{\mu\nu}\, q_\nu\,.
\label{treeV}
\end{equation}
   The one-loop-order vertex diagrams generated by the interaction terms
of Eqs.~(\ref{LfreiA}) and (\ref{NewIntTerms})
are shown in Fig.~\ref{VertexFunction:fig}.
    For $q^2=0$, the results of the diagrams in Figs.~\ref{VertexFunction:fig} (a)
and (b) read, respectively,
\begin{eqnarray}
\label{fcoefficients}
\Gamma^\mu_{a}&=& g^2\left(\gamma^\mu\, f_{1a}+i\,\sigma^{\mu\nu}\,q_\nu\, f_{2a}\right),\nonumber\\
\Gamma^\mu_{b}&=& g\xi\left(\gamma^\mu\, f_{1b}+i\,\sigma^{\mu\nu}\,q_\nu\, f_{2b}\right).
\end{eqnarray}
   For the coefficients we obtain
\begin{eqnarray*}
f_{1a}&=&-\delta Z_a,\\
f_{1b}&=&-\delta Z_b,\\
f_{2b}&=&-2\kappa \delta Z_b,
\end{eqnarray*}
and the expression for $f_{2a}$ is given in Eq.~(\ref{f2a}) in the appendix.
   Putting the results together, the form factors at $q^2=0$ are given by
\begin{eqnarray}
F_1(0) & = & 1\,,\nonumber\\
F_2(0) & = & 2\kappa +g^2(2\kappa \delta Z_a+f_{2a})\nonumber\\
&=& 2 \kappa
+\frac{g^2}{32 \pi^2 M_R \left(M_R^2+2 M_R M_N+M_N^2-M^2\right)}
[{\cal A}+{\cal B}+{\cal C}+{\cal D}]\,,
\label{ffactrors}
\end{eqnarray}
where
\begin{eqnarray}
\label{ABCD}
{\cal A}&=&-(1-4\kappa M_R)M^2_R-2M_R M_N-(1+4\kappa M_R)(M^2_N+M^2)\,,\nonumber\\
{\cal B}&=&\frac{2 A_0\left(M^2\right)}{M_R^2}
\left[3\kappa M_R^3+(1+2\kappa M_R)M_R M_N+(1+3\kappa M_R)(M_N^2-M^2)\right],\nonumber\\
{\cal C}&=&\frac{2 A_0\left(M^2_N\right)}{M_R^2}
\left[\frac{M_R^3}{M_N}+(1+\kappa M_R)M_R^2-(1+2\kappa M_R)M_R M_N
-(1+3\kappa M_R)(M_N^2-M^2)\right],\nonumber\\
{\cal D}&=&-\frac{2 B_0\left(M_R^2,M^2,M_N^2\right)}{M_R^2}
\left[\kappa M_R^5+(1+2\kappa M_R)M_R M_N(M_R^2+M_R M_N-M_N^2+M^2)\right.\nonumber\\
&&\quad\quad\quad\quad\quad\quad\quad\quad\quad\quad\left.
+2\kappa M_R^3 M^2-(1+3\kappa M_R)(M_N^2-M^2)^2
\right].
\end{eqnarray}
   As expected the charge does not get renormalized, i.e.~$F_1(0)=1$.
   Moreover, the magnetic moment does not depend on the field-redefinition parameter $\xi$.
   This is the case, because in both charge and magnetic moment
the $\xi$-dependent part of the residue of
the propagator exactly cancels the $\xi$-dependent parts of the loop vertex diagrams of
Fig.\ \ref{VertexFunction:fig} (b).
   For an unstable heavy fermion ($M_R>M_N+M$),
the latter also contain imaginary parts.
   Therefore, any alternative definition of the charge and the magnetic moment,
making use of a real-valued wave function renormalization constant, necessarily leads to
$\xi$-dependent and thus unphysical quantities.
   For example, let us denote by $M_R^R$ the mass of the heavy fermion defined
as the zero of the real part of the inverse propagator, i.e.
\begin{equation}
M_R^R - M_R -{\rm Re}\,\Sigma(M_R^R)=0\,. \label{Rmassequation}
\end{equation}
   In Refs.~\cite{Sirlin:1991fd,Sirlin:1991rt,%
Willenbrock:1991hu,Gegelia:1992kj,Gambino:1999ai,Djukanovic:2007bw}
it was shown that this definition leads to gauge- and
field-redefinition-parameter-dependent masses of the unstable particles starting at
two-loop order.
   Even though from a phenomenological point of view one might argue that a
deviation at the two-loop level is small, as a matter of principle, a physical quantity
by definition should be gauge and field-redefinition independent.
   Moreover, as we will see below, we encounter similar problems already at
one-loop order and there is no good reason to ignore this issue.

    Close to $p\hspace{-.45 em}/\hspace{.1em}\sim M_R^R $, the dressed
propagator can be written as
\begin{eqnarray}
S_R(p) &=& \frac{1}{\left(p\hspace{-.45em}/\hspace{.1em}-M_R^R\right)
\left[1-\Sigma'(M_R^R)\right]-i\,{\rm Im}\, \Sigma(M_R^R)
+{\cal O}\left[\left(p\hspace{-.45em}/\hspace{.1em}-M_R^R\right)^2\right]}\nonumber\\
&=&\frac{Z_R}{\left(p\hspace{-.45em}/\hspace{.1em}-M_R^R\right)
\left[1 -i Z_R\, {\rm Im}\, \Sigma'(M_R^R)\right]
-i\,Z_R\, {\rm Im}\, \Sigma(M_R^R)
+{\cal O}\left[\left(p\hspace{-.45em}/\hspace{.1em}-M_R^R\right)^2\right]}\,,
\label{dressedDprReal}
\end{eqnarray}
where
\begin{displaymath}
Z_R=\frac{1}{1-{\rm Re }\,\Sigma'(M_R^R)}\,.
\end{displaymath}
Up to one-loop accuracy Eq.~(\ref{dressedDprReal}) can be written
as
\begin{eqnarray}
S_R(p)&=&\frac{Z_R}{\left(p\hspace{-.45
em}/\hspace{.1em}-M_R^R\right)\left[1 -i\,{\rm Im}\,
\Sigma'(M_R^R)\right]-i\,{\rm Im}\, \Sigma(M_R^R)
+{\cal O}\left[\left(p\hspace{-.45em}/\hspace{.1em}-M_R^R\right)^2\right]}
\,\label{dressedDprRealOneloop}
\end{eqnarray}
with
\begin{displaymath}
Z_R=1+{\rm Re}\,\Sigma'(M_R^R)\,.
\end{displaymath}
Expanding the fraction and keeping only terms up to one-loop accuracy,
we obtain
\begin{displaymath}
S_R(p)=
Z_R \frac{p\hspace{-.45em}/\hspace{.1em}+M_R^R+i\,{\rm Im}\, \Sigma(M_R^R)}
{p^2-{M_R^R}^2+iM_R^R\, \Gamma_R(p^2)}
\end{displaymath}
which has the characteristic Breit-Wigner form with an
energy-dependent width
\begin{displaymath}
\Gamma_R(p^2)=-2\, {\rm Im}\, \Sigma(M_R^R)-\left(p^2-{M_R^R}^2\right)
\frac{{\rm Im}\,\Sigma'(M_R^R)}{M_R^R},
\end{displaymath}
and a real wave function renormalization constant $Z_R$
\cite{Sirlin:1991fd}.

   Using the Dirac spinors of Eqs.~(\ref{spinorsdef}) with the mass
$M_R^R$ instead of a complex $z$, putting the external legs of
the vertex functions ''on mass shell'', i.e.~$ p\hspace{-.45
em}/\hspace{.1em}=M_R^R$, and taking $Z_R$ as the wave function
renormalization constant, we obtain for the form factors the
following results:
\begin{eqnarray}
F_1(0) & = & 1+ \frac{i\,g^2\, {\rm Im} \left[B_0
\left(M_R^2,M^2,M_N^2\right)\right]}{32  \pi ^2 M_R^2
\left(M_R^2+2M_R M_N+M_N^2-M^2\right)} \biggl[M_R^4+2 M_R^3M_N\nonumber\\
&& +2 M_R^2 \left(M_N^2+M^2\right)  -2M_R M_N \left(M_N^2-M^2\right)
-3\left(M_N^2-M^2\right)^2\biggr]\nonumber\\
&& -\frac{i\,g\,\xi\,{\rm Im}\left[
B_0\left(M_R^2,M^2,M_N^2\right)\right]
\left[M_R^4-2M_R^2\left(M_N^2+M^2\right)+\left(M_N^2-M^2\right)^2\right]}
{16 \pi^2 M_R\left(M_R^2+2M_R M_N+M_N^2-M^2\right)}\,,\nonumber\\
F_2(0) & = & 2\,\kappa +\frac{g^2}{32 \pi^2  M_R
   \left(M_R^2+2M_R M_N+M_N^2-M^2\right)}\,({\cal A}+{\cal B}+{\cal C}+{\cal D}_1
   +\mbox{Re}[{\cal D}_2])\nonumber\\
&& - \frac{i\,g\,\xi\,\kappa\,
{\rm Im} \left[ B_0\left(M_R^2,M^2,M_N^2\right)\right]
\left(M_R^2-2M_R M_N+M_N^2-M^2\right)}{8\pi^2M_R}\,,
\label{ffactrorsRealZ}
\end{eqnarray}
where ${\cal A}$, ${\cal B}$, ${\cal C}$, and ${\cal D}$ are given
in Eqs.~(\ref{ABCD}), and ${\cal D}_1$ and ${\cal D}_2$ refer to the
$\kappa$-independent and $\kappa$-dependent parts of ${\cal D}$, respectively.
   From Eqs.~(\ref{ffactrorsRealZ}) it is clearly seen that within
the ``on-mass-shell'' scheme the charge of the unstable heavy fermion
receives ``strong'' corrections.
   Furthermore, both the charge and the magnetic moment have imaginary parts
which depend on the field-redefinition parameter $\xi$.
   Such a definition does not qualify as a physical observable.

   One might define the vertex function using Eq.~(\ref{FF}) with a
Breit-Wigner mass $M_R^R$ instead of $z$, in combination with the
complex wave function renormalization constant
\begin{displaymath}
Z=\frac{1}{1-\Sigma'(M_R^R)}\,.
\end{displaymath}
   At one-loop order such a definition leads to the non-renormalization of
the charge and a $\xi $-independent magnetic moment.
   This is the case because the so obtained one-loop-order result coincides with
the one extracted at the pole position.
   Now the problem shows up starting at two-loop order.

   Let us analyze the $\xi$ dependence at two-loop order.
   For simplicity we take $\kappa=0$ and consider the vertex function at
$q^2=0$ and  $p_i^2=p_f^2=z^2$ with $z$ corresponding to either the pole or
to the Breit-Wigner mass,
\begin{equation}
G^{\mu,ij}(p_f,p_i) = Z(z)\,\bar w^i(p_f) \Gamma^\mu(p_f,p_i) w^j(p_i)=Z(z)\, \bar
w^i(p_f)\left[\,\gamma^\mu\, W_1 (z) + i \,\sigma^{\mu\nu}\,
q_\nu\,W_2(z)\,\right]w^j(p_i) \,, \label{FFpar}
\end{equation}
where
\begin{equation}
Z(z)=\frac{1}{1-\Sigma'(z)}=1+\hbar\, \delta Z^{(1)}(z)
+\hbar^2\, \delta Z^{(2)}(z)+{\cal O}(\hbar^3)\,.
\label{Zloop}
\end{equation}
   In Eq.~(\ref{Zloop}), the expansion in $\hbar$ corresponds to the loop expansion.
   The Ward identity guarantees that
\begin{equation}
Z(z)\,W_1(z)=1 \label{wi},
\end{equation}
and therefore the charge is not renormalized.
   Expanding $W_2(z)$ in the number of loops,
\begin{equation}
W_2(z) = \hbar\,W_2^{(1)}(z)+\hbar^2\,W_2^{(2)}(z)+{\cal O}(\hbar^3)\,,
\nonumber \\
\label{w2loop}
\end{equation}
and substituting Eq.~(\ref{Zloop}) and $ z = M_R+\hbar\, \delta M
+{\cal O}(\hbar^2)$ into Eq.~(\ref{FFpar}),
 we obtain for the vertex function up to and including order $\hbar^2$,
\begin{eqnarray}
G^{\mu,ij}(p_f,p_i)& = & \bar w^i(p_f)\biggl(\gamma^\mu
+ i\,\sigma^{\mu\nu}\,q_\nu\,
\biggl\{\hbar W_2^{(1)}(M_R)\nonumber\\
&& +\hbar^2 \,\left[ W_2^{(2)}(M_R)+\delta
Z_1(M_R)W_2^{(1)}(M_R)+\delta M \,W_2^{(1) '}(M_R) \right]
\biggr\}\,\biggr)w^j(p_i) \,, \label{FFpar2loop}
\end{eqnarray}
where
$$W_2^{(1) '}(M_R)=\left.\frac{d W_2^{(1)}(z)}{dz}\right|_{z=M_R}.$$
   At order $\hbar$, the vertex function is independent of $\xi$ for
any choice of $\delta M$, because $W_2^{(1)}(M_R)$ does not depend on $\xi$.
   At order $\hbar^2$, for this to happen the $\xi$-dependent parts of $W_2^{(2)}(M_R)$
and of the remaining two terms have to precisely cancel each other.
   The calculated result for $W_2^{(1)'}(M_R)$ contains a non-vanishing
term linear in $\xi$.
   For $z$ chosen as the pole position of the dressed propagator this cancelation
has to take place as the residue of the $S$-matrix cannot depend on $\xi$.
   For any other choice of $z$ (i.e.~$\delta M$), including the Breit-Wigner mass, a
$\xi$ dependence remains at two-loop order.
   We therefore conclude that the physical quantities characterizing the unstable
particles should be defined at the complex pole of the dressed propagator, because
only this specification guarantees a $\xi$-independent result at arbitrary
loop order.

\section{Summary}
   In the framework of effective quantum field theory we have addressed
the definition of physical quantities characterizing unstable particles.
   To that end we considered a charged, unstable heavy fermion
which we allowed to decay into a charged, stable light fermion
and a stable, neutral pseudoscalar.
   While our arguments are applicable to all physical quantities
characterizing unstable particles, we have focussed on the charge and
the magnetic moment of the resonance.
   Our discussion made use of the well-known fact that observables should
remain invariant under a field transformation.
   With this in mind, we performed in our model Lagrangian a particular
field transformation depending on an arbitrary parameter $\xi$.
   By appealing to the $\xi$ independence of physical quantities,
in a one-loop calculation we demonstrated that physical properties
characterizing unstable resonances should be defined through the
residues of the $S$-matrix in the complex plane.
   As opposed to this, if the ''on-mass-shell'' scheme is used then even
the charge and not only the magnetic moment
will receive contributions from the strong interactions and both
will depend on the parameter $\xi$.
   In summary, the main conclusion is that physical quantities
characterizing unstable particles should be extracted from the
residues at complex poles.
   This observation neither depends on the choice of
observables nor on the details specific to the model.

\acknowledgments

   The authors thank D.~Djukanovic for providing the computer
programs for the calculation of Feynman diagrams and for comments
on the manuscript.
   This work was supported by the Deutsche Forschungsgemeinschaft
(SFB 443).

\section{Appendix}
In the following we collect the contributions of the loop diagrams
of Figs.~\ref{DeltaMassInd:fig} and \ref{VertexFunction:fig} to
the wave function renormalization constant and the vertex functions,
respectively.
   The loop functions $A_0$ and $B_0$ are defined as
\begin{eqnarray}
A_0(m^2)& = & \frac{(2 \pi)^{4-n}}{i\,\pi^2}\,\int \frac{
d^nk}{k^2-m^2+i\,0^+}\,, \nonumber\\
B_0(p^2,m_1^2,m_2^2)& = & \frac{(2
\pi)^{4-n}}{i\,\pi^2} \,\int\frac{d^nk
}{\left[k^2-m_1^2+i\,0^+\right]\left[(p+k)^2-m_2^2+i\,0^+\right]}\,.\nonumber
\label{oneandtwoPF}
\end{eqnarray}
   The coefficients $\delta Z_a$ and $\delta Z_b$ of Eq.\ (\ref{deltaZ}) are given by
\begin{eqnarray}
\label{deltaZa}
\delta Z_a & = &\frac{M_R^2-M_N^2-M^2}
{16 \pi ^2\left(M_R^2+2 M_R M_N+M_N^2-M^2\right)}\nonumber\\
&&+\frac{\left(3M_R^2+2M_R M_N+3M_N^2-3M^2\right)A_0\left(M^2\right)}
{32 \pi^2 M_R^2\left(M_R^2+2 M_R M_N+M_N^2-M^2\right)}\nonumber\\
&& +\frac{\left(M_R^2-2M_R M_N-3 M_N^2+3 M^2\right)A_0\left(M_N^2\right)}
{32 \pi ^2 M_R^2 \left(M_R^2+2 M_R M_N+M_N^2-M^2\right)}\nonumber\\
&& -\frac{B_0\left(M_R^2,M^2,M_N^2\right)}{32 \pi^2 M_R^2
\left(M_R^2+2 M_R M_N+M_N^2-M^2\right)}\nonumber\\
&& \times \biggl[M_R^4+2M_R^3M_N+2M_R^2M_N^2+2M_R^2M^2-2M_R M_N^3
\nonumber\\
&&\,\,\,\,\,\,\,+2 M_R M_N M^2-3M_N^4+6M_N^2M^2-3M^4\biggr]
\end{eqnarray}
and
\begin{equation}
\label{deltaZb}
\delta Z_b=
\frac{1}{16\pi^2 M_R}
\biggl\{A_0\left(M^2\right)-A_0\left(M_N^2\right)
+\biggl[M_R^2+M_N^2-2 M_R M_N-M^2\biggr]B_0\left(M_R^ 2,M^2,M_N^2\right)\biggr\} \,.
\end{equation}
   The coefficients $f_{1a}$, $f_{2a}$, $f_{1b}$, and $f_{2b}$
of Eqs.\ (\ref{fcoefficients}) are given by
\begin{eqnarray}
f_{1a}&=&-\delta Z_a,\nonumber\\
f_{2a}&=&\frac{1}{32 \pi ^2  M_R^3 M_N\left(M_R^2+2 M_R M_N+M_N^2-M^2\right)}\nonumber\\
&&\times\biggl\{-M_R^2 M_N\left(M_R^2+2M_R M_N+M_N^2+M^2\right) \nonumber\\
&&\,\,\,\,\,\,\,\,\,+ 2  M_N \left(M_R M_N+M_N^2-M^2\right)A_0\left(M^2\right)\nonumber\\
&&\,\,\,\,\,\,\,\,\,+2\left(M_R^3+M_R^2 M_N-M_R M_N^2-M_N^3+M_N M^2\right)A_0\left(M_N^2\right)\nonumber\\
&&\,\,\,\,\,\,\,\,\,-2 M_N \left(M_R^3 M_N+M_R^2 M_N^2-M_R M_N^3+M_R M_N M^2-M_N^4+2M_N^2M^2-M^4\right)
\nonumber\\
&&\,\,\,\,\,\,\,\,\,\times B_0\left(M_R^2,M^2,M_N^2\right)\biggr\}\,,\label{f2a}\\
f_{1b}&=&-\delta Z_b,\nonumber\\
f_{2b}&=&-2\kappa\delta Z_b\nonumber.
\end{eqnarray}

\end{document}